\documentclass[%
 reprint,
superscriptaddress,
showpacs,
showkeys,
 amsmath,amssymb,
 aps,
prb,
]{revtex4-2}
\usepackage{graphicx,amssymb,amsmath,epsfig}
\usepackage{color}
\usepackage{comment}

\usepackage{mhchem}

\usepackage{hyperref}
\hypersetup{
    colorlinks=true,
    linkcolor=blue,
    filecolor=blue,      
    urlcolor=blue,
    citecolor=blue,
    pdftitle={Overleaf Example},
    pdfpagemode=FullScreen,
    }

\begin{document}

\title{Tuning the Electronic and Optical Properties of Two-Dimensional Diboron-Porphyrin by Strain Engineering: A Density Functional Theory Investigation}
\author{Isaac M. Felix}
\affiliation{Departament of Theoretical and Experimental Physics, Federal University of Rio Grande do Norte, Natal, RN, 59072-970, Brazil.}
\author{Fabiano M. Andrade}
\affiliation{Department of Mathematics and Statistics, State University of Ponta Grossa, Ponta Grossa, PR, 84030-000, Brazil.}
\author{Cristiano F. Woellner}
\affiliation{Physics Department, Federal University of Paran\'a, UFPR, Curitiba, PR, 81531-980, Brazil}
\author{Douglas S. Galv\~ao}
\affiliation{Applied Physics Department, University of Campinas, Campinas, S\~ao Paulo, Brazil.}
\affiliation{Center for Computing in Engineering and Sciences, University of Campinas, Campinas, S\~ao Paulo, Brazil.}
\author{Raphael M. Tromer}
\affiliation{School of Engineering, Mackenzie Presbyterian University, São Paulo SP. 01302-907, Brazil}
\affiliation{MackGraphe — Mackenzie Institute for Research in Graphene and Nanotechnologies, Mackenzie Presbyterian Institute, São Paulo SP. 01302-907, Brazil}
\affiliation{Applied Physics Department, University of Campinas, Campinas, S\~ao Paulo, Brazil.}
\affiliation{Center for Computing in Engineering and Sciences, University of Campinas, Campinas, S\~ao Paulo, Brazil.}
\date{\today}

\begin{abstract}
In the present work, we have carried out DFT simulations to investigate the electronic and optical properties of a porphyrin-based 2D crystal named 2D Diboron-Porphyrin (2DDP). We showed that it is possible to use strain to tune the 2DDP electronic properties (from semiconductor to metal) depending on the direction of the applied strain. 2DDP exhibits optical activity from the infrared to the ultraviolet region. Similarly to electronic bands, strain can also modulate the optical activity response. 2DDP can be a promising candidate for some electro-opto-mechanical applications.

\end{abstract}



\maketitle

\section{Introduction}

The quest for novel 2D materials with tailored properties has led to significant advances in nanotechnology and materials science. Among these materials, graphene, isolated from bulk graphite through a mechanical exfoliation method in 2004, stands as a prominent example \cite{Novoselov2004,Novoselov2005}. 

Graphene's remarkable electronic, mechanical, and thermal properties have been exploited in various technological applications, especially in flexible electronic devices \cite{Gwon2011}. However, despite graphene's exceptional attributes, its intrinsic semi-metallic nature (zero electronic band gap) presents a challenge for its incorporation into nanoelectronic devices that rely on controlling on-off current behavior \cite{Xia2010}. To address this limitation and unlock the full potential of 2D materials for nanoelectronics, researchers have explored strategies to engineer band gaps in these materials \cite{Cadelano2010}. One possibility along these lines is the use of porphyrin-based structures. Due to their versatile chemical nature, porphyrins are excellent building blocks for creating large 2D systems \cite{Tan2013,He2017,Ramirez2017}.

These porphyrin-based systems have attracted great interest for their potential applications in diverse fields, ranging from hydrogen storage to catalysis and spintronics \cite{Lee2011,Ali2009,Cho2011}. Moreover, their ability to form polymers opens opportunities for creating extensive 2D networks with tailored properties \cite{Abel2010}. Among these structures, it is worth mentioning a porphyrin-based structure with two boron atoms at its core, forming a \ce{B2} bond \cite{Brothers2011}. The resulting diboranyl-porphyrins have been synthesized and extensively studied \cite{Arnold1990,Brand1994,Weiss2007,Belcher2008}.

Building on the knowledge from prior research on porphyrin-based materials and the interesting properties of diboranyl-porphyrins and inspired by a structure investigated by P.J. Brothers et al. \cite{Brothers2011}, we propose recently a new 2D semiconductor material, named two-dimensional diboron-porphyrin (2DDP) \cite{Tromer2020}. In this work, we consider a unit cell composed of a diboron-porphyrin molecule, wherein the usual C-H bonds are replaced by C–C covalent bonds, leading to the formation of a 2D supercell with \ce{BC10N2} stoichiometry (see Figure \ref{fig:bands}). 

We carried out a comprehensive computational investigation using first-principles methods to assess 2DDP's structural stability and physico-chemical properties. We investigated its mechanical stability and electronic properties. Its electronic band structure reveals 2DDP to be a direct-gap semiconductor with an electronic band gap of 0.6 eV. The results show an anisotropic behavior along the in-plane direction. Finally, we investigated the behavior of the optical properties as a function of an externally applied strain. 

Our results show that the proposed 2DDP exhibits semiconductor properties suitable for optoelectronics applications. The information and insights gained from this work could help to design and develop innovative novel 2D materials with tailored properties.

\section{Methodology}

We carried out DFT calculations using the SIESTA code\cite{Soler2002} to investigate the 2DDP electronic and optical properties\cite{Tromer2020}. The simulations were performed in generalized gradient approximation (GGA). The core electrons were described using the norm-conserving Troullier-Martins pseudopotential \cite{Troullier1991}. We used the basis sets double-zeta plus polarization (DZP). 
A kinetic energy cut-off of 300 eV was employed. We used a $6\times 6\times 1$ k-grid for geometry optimization, while electronic and optical calculations were performed on a denser $30\times 30\times 1$ k-grid. To prevent spurious interactions among the 2D layers, a vacuum region of 20 \AA\ was included.
The lattice vectors and atom positions were fully relaxed during the optimization process, and we set convergence criteria for the maximum force on each atom ($< 1.0 \times 10^{-3}$ eV/\AA) and the total energy ($< 1.0 \times 10^{-6}$ eV). In a previous study \cite{Tromer2020}, we investigated the stability of the 2DDP. The analysis used phonon dispersion calculations at $T = 0$ K and \textit{ab initio} molecular dynamics (AIMD) simulations at 4000 K. 

These complementary approaches allowed us to assess the material’s stability at low and elevated temperatures. We applied an external electric field of 1.0 V/\AA\ along the $x$ and $y$ directions to determine the optical properties. 
This field strength has been used in the literature for other 2D systems similar to 2DDP, such as nitrogenated holey graphene\cite{Mahmood2015} and its derivatives doped with boron and nitrogen \cite{Tromer2020pccp}. The optical quantities were derived directly from the complex dielectric constant $\epsilon = \epsilon_1 + i\epsilon_2$ where $\epsilon_1$ and $\epsilon_2$ represent the real and imaginary parts of the dielectric constant, respectively. Using Fermi’s golden rule, we obtained the imaginary part of the dielectric constant by considering interband optical transitions between valence (VB) and conduction bands (CB):

\begin{equation}
\epsilon_2(\omega)=\frac{4\pi^2}{V_\Omega\omega^2}\displaystyle\sum_{i\in \mathrm{VB}, \, j\in \mathrm{CB}}\displaystyle\sum_{k} W_k \, |\rho_{ij}|^2 \, \delta	(\varepsilon_{kj}-\varepsilon_{ki}- \hbar \omega),
\end{equation}
where $\omega$ is the photon frequency, $V_\Omega$ the unit cell volume, $W_k$ the k-point weight in the reciprocal space, and $\rho_{ij}$ the dipole transition matrix element \cite{Tromer2021}. $\epsilon_1$ and $\epsilon_2$ are related through the Kramers-Kronig relation. The real part of the dielectric constant is expressed as:

\begin{equation}
\epsilon_1(\omega)=1+\frac{1}{\pi}P\displaystyle\int_{0}^{\infty}d\omega'\frac{\omega'\epsilon_2(\omega')}{\omega'^2-\omega^2},
\end{equation}
where P is the principal value of the logarithm. Once we obtain the real and imaginary parts of the dielectric constant, denoted as $\epsilon_1$ and $\epsilon_2$, we can calculate other important optical coefficients. These coefficients include the absorption coefficient $\alpha$, the refractive index $\eta$, and the reflectivity $R$. These quantities can be derived using the following expressions:

\begin{equation}
\alpha (\omega )=\sqrt{2}\omega\bigg[(\epsilon_1^2(\omega)+\epsilon_2^2(\omega))^{1/2}-\epsilon_1(\omega)\bigg ]^{1/2},
\end{equation}
\begin{equation}
\eta(\omega)= \frac{1}{\sqrt{2}} \bigg [(\epsilon_1^2(\omega)+\epsilon_2^2(\omega))^{1/2}+\epsilon_1(\omega)\bigg ]^{2},
\end{equation}
and
\begin{equation}
R(\omega)=\bigg [\frac{(\epsilon_1(\omega)+i\epsilon_2(\omega))^{1/2}-1}{(\epsilon_1(\omega)+i\epsilon_2(\omega))^{1/2}+1}\bigg ]^2.
\end{equation}

\section{Results}

\begin{figure}[t]
    \centering
    \includegraphics[width=0.8\linewidth]{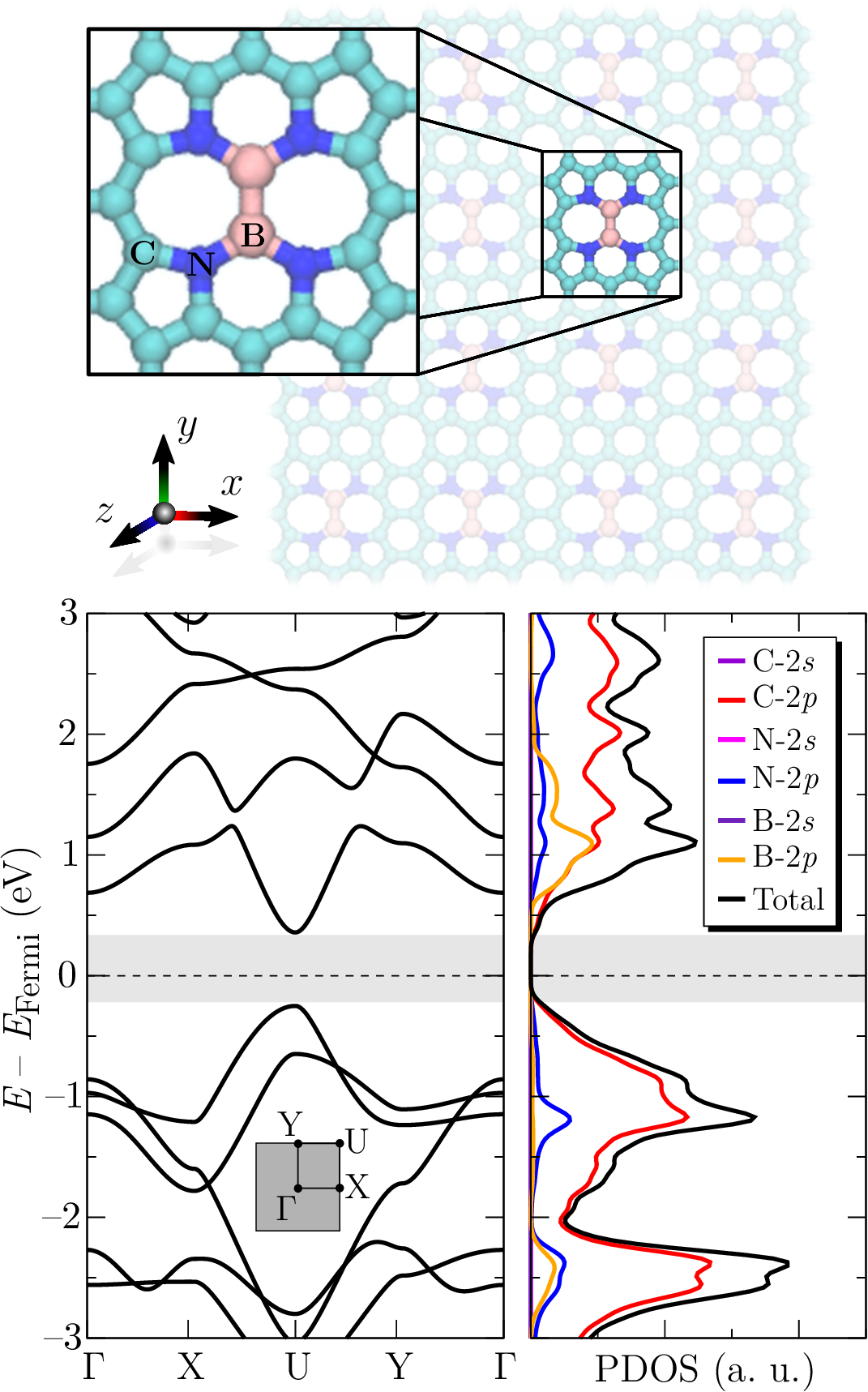}
    \caption{Top: The used unit cell; Botton: Electronic band structure calculated along $\Gamma \rightarrow~\mathrm{X}~\rightarrow~\mathrm{U}~\rightarrow~\mathrm{Y}~\rightarrow~\Gamma$ path and the corresponding projected density of states (PDOS).}
    \label{fig:bands}
\end{figure}

To perform our analysis, first, we reproduced the parameters obtained in the previous investigation for 2DDP shown in Figure \ref{fig:bands}) \cite{Tromer2020}.
We obtained the following lattice parameters for the optimized 2DDP structure: $a_{x}=8.19$~\AA, $b_{y}=8.69$~\AA, $c_z=20.00$~\AA~(vacuum region), $\alpha=\beta=\gamma=90^o$, and, the relevant bond distances are $R_{BB}=1.67$~\AA, $R_{BN}=1.47$~\AA, $R_{CN}=1.44$~\AA~and, $R_{CC}$ from $1.42$ to $1.45$ \AA.

In Figure \ref{fig:bands}, we present the 2DDP electronic band structure calculated along $\Gamma \rightarrow~\mathrm{X}~\rightarrow~\mathrm{U}~\rightarrow~\mathrm{Y}~\rightarrow~\Gamma$ special lattice points and the corresponding projected density of valence states (PDOS).  This Figure shows that 2DDP is a semiconductor with a direct band gap at $\mathrm{U}$ point, with a value around 0.6 eV, consistent with previous work \cite{Tromer2020}. The electronic band gap value is expected to be underestimated as we used a GGA-PBE functional \cite{Perdew1996}. A more precise band gap calculation using HSE-06 functional \cite{Krukau2006} produced a value around 1.0 eV \cite{Tromer2020}, confirming the semiconductor characteristics. 

As mentioned above, this work focused on investigating the changes in electronic and optical properties when an external strain is applied. Understanding these changes is paramount for many applications. 

From PDOS in Figure \ref{fig:bands}, we can see that the top valence states (or the highest occupied crystalline orbital (HOCO)) are essentially composed of the $2p$ states from carbon atoms. We can also notice a slight contribution from nitrogen from the $2p$ atomic orbital. In contrast, the conduction bottom states (or lowest unoccupied crystalline orbital (LUCO)), besides the contribution of $2p$ carbon, also exhibit small contributions from nitrogen and boron atoms. It follows the expected chemical trends where boron atoms have donor characteristics. We also observe from PDOS that the $s$ states slightly contribute to the HOCO and LUCO, which are dominated by $p$ states from carbon, nitrogen, and boron atoms.
 
\begin{figure}[htbp]
    \centering
    \includegraphics[width=\linewidth]{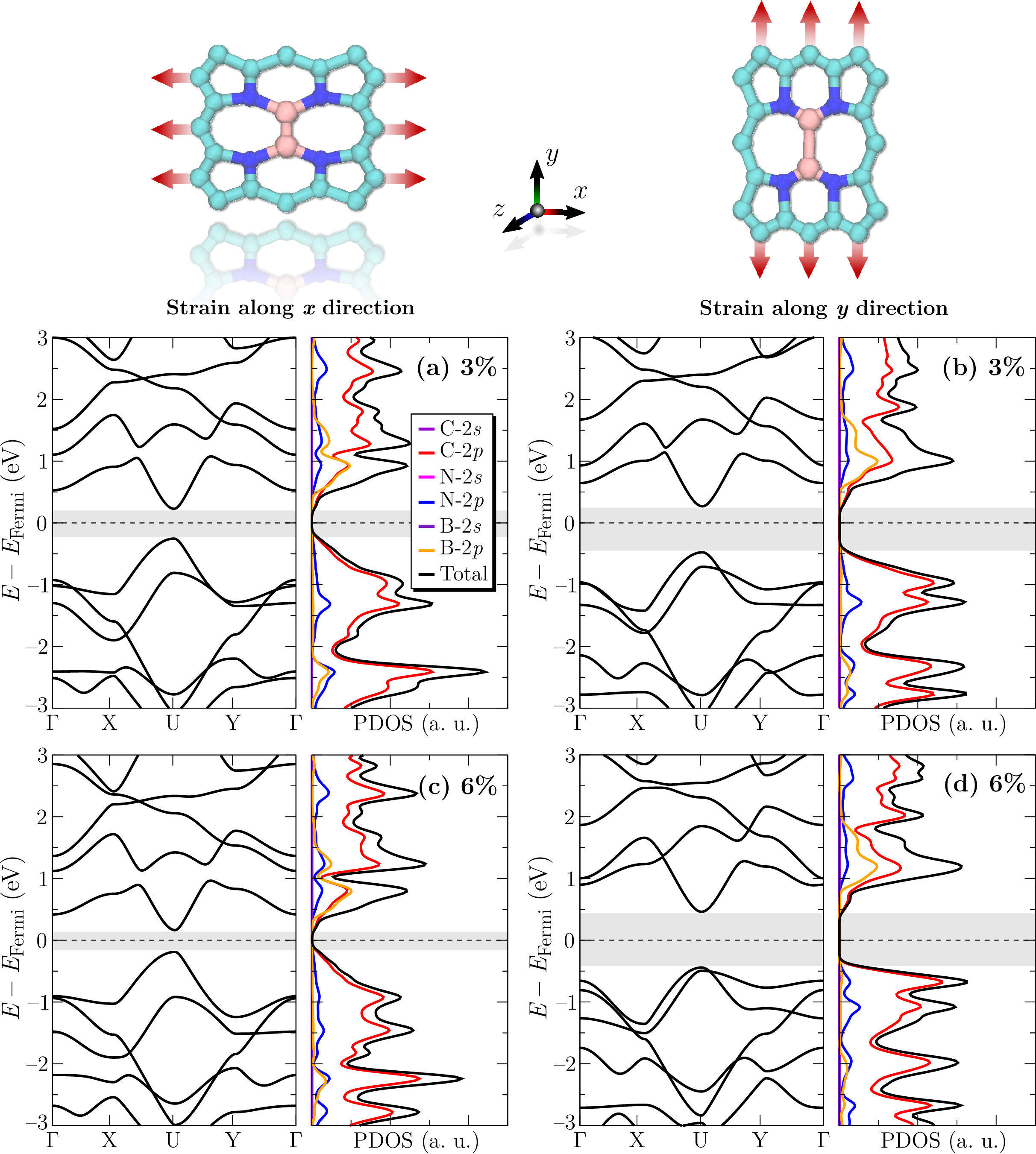}
    \caption{Top: Directions of the externally uniaxial applied strain. Botton: Electronic band structure calculation along the $\Gamma \rightarrow~\mathrm{X}~\rightarrow~\mathrm{U}~\rightarrow~\mathrm{Y}~\rightarrow~\Gamma$ path and the corresponding projected density of states (PDOS) for strain values of 3 and 6\%, for the x (left) and y (right) directions.}
    \label{fig:bands-strain}
\end{figure}

In Figure \ref{fig:bands-strain}, we present the electronic band structure calculation results for the cases of 3 and 6\% uniaxial strain applied along the 2DDP $x$ and $y$ directions, respectively. From the bands and corresponding partial density of states (PDOS), we observe that the strain causes a decrease in the band gap value when applied along the $x$ direction, while an increase occurs for the $y$ direction. In contrast, we recently reported that hydrogen atom/molecule adsorption on 2D metallic porphyrin systems were not as sensitive to the application of uniaxial strain \cite{Tromer2024}.

The distinct behavior for the $x$ and $y$ directions can be explained in terms of the relative orientation of the B-B bonds. 2DDP is not auxetic; its dimensions are reduced along the direction perpendicular to the applied strain. When 2DDP is stretched along the $x$ direction, the applied strain is perpendicular to the B-B bond direction. In this case, the B-B bond is shortened, which results in an energy red shift of the band gap value. Increasing the strain to 6 \% further increases this effect. The HOCO and LUCO energy difference becomes zero at the symmetry point U, causing 2DDP to acquire metallic characteristics.

On the other hand, when the strain is applied along the $y$ direction, the strain is parallel to the B-B bond direction, causing it to stretch along the same direction as the applied strain. In this case, it produces a blue shift in the band gap values, which also increases when the strain is increased, thus reinforcing the semiconductor behavior. Therefore, depending on the direction of the externally applied strain and its value, it is possible to tune the 2DDP electronic behavior (increasing or decreasing the semiconductor behavior or even inducing a metallic one.) Electronic band gap tuning is of fundamental importance for some electro-mechanical applications.

Once we observed that the electronic properties of 2DDP can be tuned by applying a uniaxial strain, inducing changes in its electronic states, we extended our analyses to determine the strain effects on the 2DDP optical properties when interacting with electromagnetic radiation. We have considered the photon energy range to be from 0 to 8 eV, corresponding to the spectrum from infrared to ultraviolet.

\begin{figure}[htbp]
    \centering
    \includegraphics[width=\linewidth]{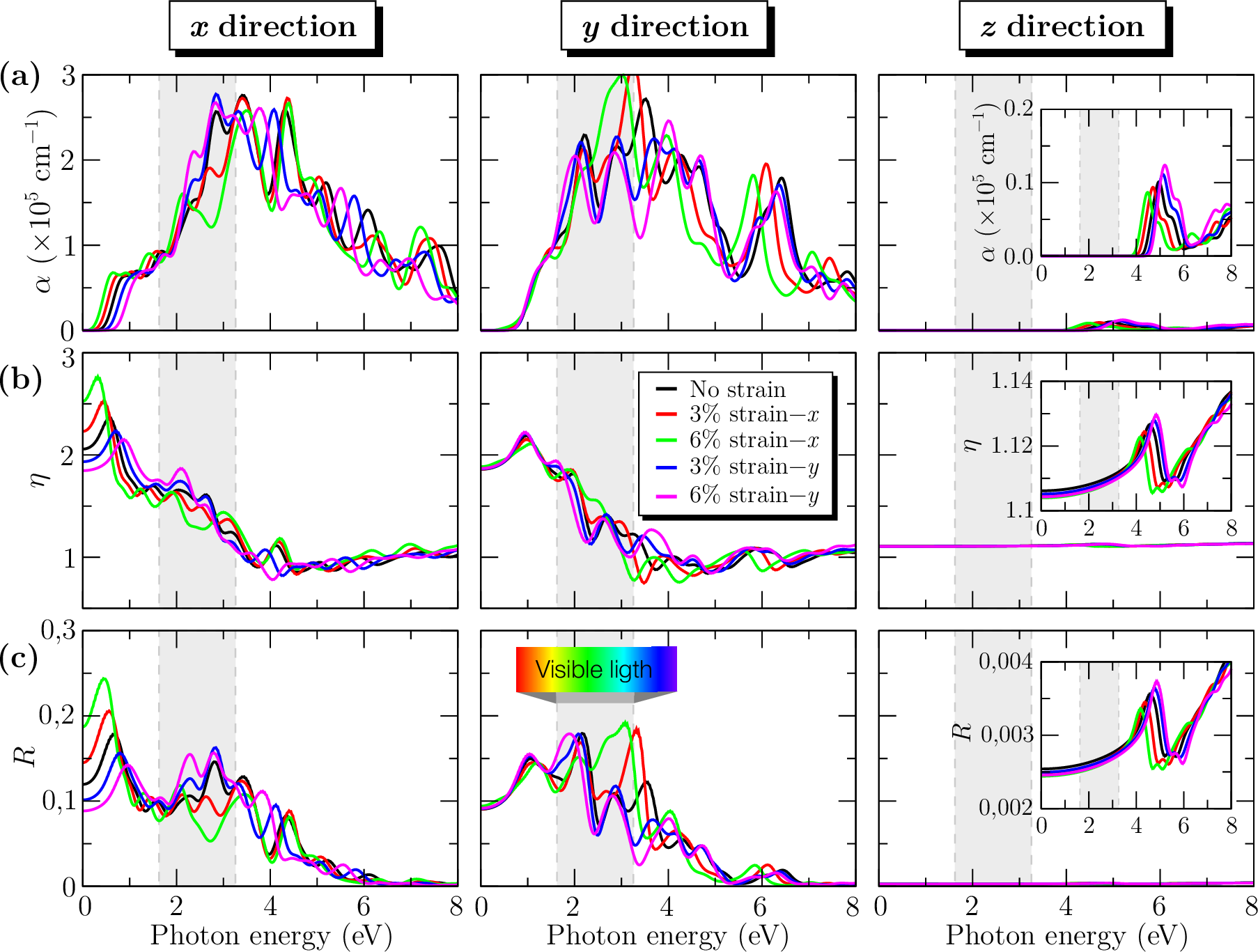}
    \caption{2DDP optical properties coefficients of the absorption coefficient $\alpha$ (a), refractive index $\eta$ (b), and the reflectivity $R$ (c) as a function of both photon energy along $x,y,z$, and the applied strain values.}
    \label{fig:optic}
\end{figure}

The corresponding results for the optical properties, obtained using the expressions outlined in the methodology section, are presented in Figure \ref{fig:optic}. Each column in Figure \ref{fig:optic} represents a specific direction corresponding to the polarization of the incident photon beam. As we can observe from this Figure, the interaction for the $z$-axis is significantly weaker than for the other two directions. This is a consequence of the 2D topology and prevents many optoelectronic applications such as data storage or high-performance computing \cite{Youngblood2016}.

For the $x$ and $y$ directions, 2DDP exhibits an optical response for almost the entire analyzed spectrum. Regarding the absorption coefficient, a series of peak absorptions is present in the visible-ultraviolet interface. Closely examining the corresponding black curves indicates a slight anisotropy for the $x$ and $y$ directions. The optical absorption starts near $0.6$ eV for both directions, which is consistent with the direct gap at the symmetry point U, where the optical transitions involve the localized crystalline orbitals and the HOCO and LUCO ones, characterizing a $\pi$ $\rightarrow$ $\pi^*$ transition.

The intensity of absorption peaks is smaller in the infrared region and at the far edge of the ultraviolet one. In the visible region, some peaks are in the green region, nearly at the intersection with the ultraviolet. When the strain is applied, as observed in the cases of the isolated structures discussed above, the band gap values change depending on the direction of the applied strain. Since optical transitions strongly depend on the orbitals' values, they undoubtedly change under the strain regime. We verified this effect on the absorption plots, where there is a clear red shift when the strain is applied along the $x$ direction and a blue one when applied along the $y$ direction. These shifts increase when the strain is increased from 3 to 6\%. Also, the strain effects are less pronounced for light polarization along the $y$ direction than for $x$, where the shifts in optical transitions are more evident. This behavior was also observed for 2D-TTA \cite{Tromer2021}.

In Figure \ref{fig:optic}, we present the refractive (\ref{fig:optic}-b) and reflectivity  (\ref{fig:optic}-c) results for the $x$ and $y$ polarizations. The infrared region exhibits a notable peak in reflectivity, coinciding with a significant absorption, around energy values of 1.25--1.50 eV under strain regimes. Simultaneously, the refractive index increases in this region. With a maximum of 25\% of infrared light being reflected. The high refractive index suggests that most of the incident light is deflected towards refraction. Consequently, when light is shone on 2DDP, most penetrates it, with only a small fraction being reflected.

This trend extends to the visible and ultraviolet regions. In the visible range, the refractive index surpasses reflectivity by at least one order of magnitude. This distinction is important for evaluating the material's suitability in solar cells and optical device applications \cite{Yang2014,Bhattacharya2015}. In the ultraviolet region, reflection is minimal, below 10\%, indicating that the incident light is refracted. In the limit of small photon energy (tending to zero), we can estimate the 2DDP static refractive index of (\ref{fig:optic}-b), first and second panels). The estimated numbers are $2.1$ and $1.8$ for the $x$ and $y$ directions, respectively. These results suggest that 2DDP could be a good candidate for some optoelectronic devices, exhibiting significant absorption from $1.5$ to $6.0$ eV across the infrared to ultraviolet spectrum.

\section{Conclusions}

In summary, we have carried out an extensive DFT investigation on 2DDP electronic and optical properties. We have also investigated the effect of an externally applied strain on these properties. We showed that it is possible to use strain to tune the 2DDP electronic properties. The electronic band gap energy for the unstrained system is approximately 0.6 eV and can be increased when the strain is applied along the direction perpendicular to the B-B bond. Conversely, the system can exhibit metallic properties when the strain is applied parallel to the direction of the B-B bond. 

We have also observed that 2DDP exhibits considerable optical activity from the infrared to the ultraviolet region. Similarly to electronic bands, strain can also modulate the optical activity response, inducing red or blue shifts in the band gap values, depending on the direction of application. Therefore, the structure of 2DDP can be a promising candidate for some electro-opto-mechanical applications.

\section*{Acknowledgements}
This work was financed by the Coordenação de Aperfeiçoamento de Pessoal de Nível Superior (CAPES) -
Finance Code 001, Conselho Nacional de Desenvolvimento Científico e Tecnológico (CNPq), and
FAPESP. We thank the Center for Computing in Engineering and Sciences at Unicamp for financial support through the FAPESP/CEPID Grants \#2013/08293-7 and \#2018/11352-7.
FMA acknowledges financial support by CNPq Grant No. 313124/2023-0.
\bibliography{References}

\end{document}